# Vectorial characterization of Bloch surface wave via one-dimensional photonic-atomic structure

M. Asadolah Salmanpour, M. Mosleh, S. M. Hamidi

*Magneto-plasmonic Lab, Laser and Plasma Research Institute, Shahid Beheshti University, Tehran, Iran.*
m_hamidi@sbu.ac.ir

**Abstract:**

Use of hot atomic vapor as a new tool for tracing the complex nature of light has become a knowledge-based topic in recent years. In this paper, we examine the polarization ellipse of the Bloch surface wave (BSW) through the effect of a magnetic field on the coupling of these surface waves in BSW-hot atomic vapor cell. For this purpose, we fabricate a one-dimensional photonic crystal-based Bloch wave atom cell, where under different configurations of magnetic field, polarization ellipse of Bloch surface waves has been recorded experimentally. Our results indicate that by applying the magnetic field in different directions, Faraday and Voigt, the characteristics of electromagnetically induced transparency (EIT-like) of hybrid system change. We have used these changes to redefine the geometry of Voigt and Faraday for evanescent waves, as well as to measure the ratio of the components of the elliptical polarized electric field. These characterizations can open new insight into the miniaturized atomic field in high quality and low volumetric areas.

**Keywords:** Bloch surface wave; magnetic field; BSW-atomic vapor cell; Voigt and Faraday configurations.

## I. Introduction:

Properties of evanescent waves are considered in the design of new possibilities, especially in the optics of nanotechnology. Nanophotonics, which deals with the manipulation of light at the nanoscale, plays a significant role in enabling integrated photonics by providing novel functionalities, improved performance, and miniaturization opportunities. Electromagnetic near fields play a key role in applications of various fields, including data communication [1], optical computing [2], quantum information processing [3], biosensors [4], and biomedical imaging [5]. Many of these applications particularly rely on the transfer of energy or information between adjacent elements through near-field interactions. For example, in metasurfaces, engineering the evanescent wave coupling between neighboring individual elements in the nanostructures to achieve a wide range of functionalities, such as beam steering [6], polarization conversion [7], wavefront shaping [8], and wavefront modulation [9].

When considering near-field interactions between confined electromagnetic waves and optical elements, the polarization state of nearfields becomes particularly important [10]. When electromagnetic waves are spatially confined at the nanoscale, the polarization state of evanescent waves or propagative near fields gets a different polarization state from the exciting field. For instance, in total internal reflection of incoming transverse magnetic (TM) polarized light, in addition to transverse components of the electromagnetic field, a longitudinal field component parallel to the propagation direction emerges. This causes an elliptical polarization of the evanescent field with locked handiness to the propagation direction. Spin-momentum locking and intrinsic property of coupling between polarization states and phase gradients of evanescent fields shapes profile of non-trivial spatial structure for surface waves. topological features of surface waves make growing interest for evanescent waves in field of structured light [11-12].

Knowledge about the polarization state of evanescent waves can appear in some approaches such as surface plasmon polaritons (SPPs) at the metal-dielectric interface, Bloch surface waves (BSWs) along one-dimensional (1D) photonic crystal interfaces, and so on. The direct measurement of the polarization state of evanescent waves can be challenging due to the subwavelength localized nature of evanescent fields and decaying far from the interface. Conventional methods such as polarimetry methods, which rely on far-field, cannot be used for measuring the polarization of evanescent waves as well as their electromagnetic field intensity distribution .

To solve this bottleneck, there are some methods such as scanning near field optical microscopy (SNOM) [13], scattering by nano particles or nanostructures [14], and spectroscopic measurements [15] in order to study properties of evanescent fields. The energy of evanescent fields would be attenuated by touch of absorbing material, depending on properties of both of evanescent field and material. These phenomena would be used to directly measure the polarization state of evanescent waves [16]. Spectroscopic absorption measurement of atoms near the surface is one of the methods for direct characterization of polarization state of evanescent fields. When Atoms interact with the evanescent fields, they selectively absorb different states of polarizations based on inductive selection rules. In alkaline metal vapor by exerting an external magnetic field, degeneracy breaks and the polarization status of the electromagnetic field can be mapped on atomic transition absorptions in distinct frequency detuning. For instance, in our previous work we showed that the changes of absorption spectroscopy of atoms near the prism surface are sensitive to variations in the polarization state generated by sweep of angle of incidence of light to prism [17]. Also, there is research dealing with atomic transition selections in optogalvanic evanescent spectroscopy of Ar gas to investigate polarization changes applied to atoms near the surface in response to the alteration of polarization of incoming light [15].

BSWs are propagative evanescent electromagnetic waves tightly stocked at the interface between periodic structure and the surrounding medium. Unique features of BSWs such as long propagation distances and strong confinement to the surface of the structure, make them as basis for many low loss integrating paradigms in photonics [18-22].

Recently, we reported coupling of BSWs with atomic transitions in rubidium vapor as a paradigm to further confine interaction volume in chip-scale hot vapor-based applications [23]. Our results revealed that such coupling leads to EIT-like resonance and was tried to demonstrate that the atomic line spectroscopy can help us depict changes of local density of modes of BSW related to sweep of incidence light. Alkali metals have a wide range of applications in both fundamental and practical research such as slow and stored light [24], generation of squeezed light sources [25], atomic clocks [26], metrology [27], and quantum memories [28].

Many efforts have been made to introduce evanescent fields-based light-matter interaction as a paradigm for designing chip-scale devices based on atomic vapor cells for achieving strong light-matter interaction in new nanophotonic devices with enhanced performance in optical communications, quantum information processing, and other applications [29-31]. In this paper, noting the importance of studies that show different aspects of evanescent wave-atom couplings, we employ the magneto-optical spectroscopy in a hybrid BSW-atomic hot vapor cell to characterize the polarization state of BSWs. Further, for the first time, by analyzing measurement results in Faraday and the Voigt configurations for evanescent waves, we report the ratio between the longitudinal and the transverse components of the electromagnetic field of BSW. We believe by applying the magnetic field in three different directions relative to the direction of the transverse spin of the evanescent waves, this approach can be used for near-field vectorial imaging.

## II. Materials and methods:

A photograph of the Bloch surface wave-atom hybrid system is illustrated in Figure (1a). The hybrid system consists of one-dimensional (1D) photonic crystal composed of BK7/(SiO$_2$/ZrO$_2$)$^{12}$/SiO$_2$, which is designed and fabricated via electron gun deposition method to support a TM-polarized surface wave at wavelength according to the D1 line of rubidium (Rb We employed epoxy bonded cylindrical glass cell to the surface of photonic crystal, vacated and filled the bonded cell with natural Rb and sealed the end of glass cell through glass blowing technique to achieve a handmade BSW-atomic cell. Truncated 1D-photonic crystal can sustain BSWs, which are electromagnetic waves confined at the interface between a periodic dielectric multilayer and atomic media. BSW excites using a prism in the Kretschmann configuration. The cell is enclosed in a homemade oven and heated to a temperature of ~70 °C. The BSW-atomic cell is illuminated with a collimated laser beam at the 795 nm wavelength according to the D1 line of Rb with a transverse magnetic (TM) polarization. As we have shown in our previous article, the coupling of Bloch surface wave (BSW) and Rb atom leads to EIT-like resonance [tekrar-salmanpour article]. Generally, coupling of the atomic state as a discrete quantum state and BSW resonance as continuum states leads to Fano resonance [32]. Electromagnetically-induced transparency is a special case of Fano resonance, which occurs when frequencies of the BSW and atomic resonances match together, ($\omega_{BSW} = \omega_{atom}$). Indeed, the spectrum that comes from EIT-like phenomena is the inverse of the absorption spectrum of atomic media, which we record in our experimental method based on frequency modulation technique.

In the next step, the effect of the magnetic field (800 Gauss) on this hybrid system was investigated. The main idea is to measure the reflection spectrum of the BSW-atomic hybrid system in the Faraday and Voigt configurations, where the magnetic field vector is parallel or normal to the wavevector of the light respectively. The direction of the magnetic field determines the atomic quantization axis. In transmission spectroscopy at free space as Faraday configuration, according to the selection rules for transitions between the ground and excited hyperfine atomic states (in the dipole approximation), the atom must pass $\Delta m_F = \pm 1$ which associated with $\sigma^+$ and $\sigma^-$ transitions. Further, in Voigt configuration, the selection rules induce $\Delta m_F = 0, \pm 1$ which is associated $\pi$ and $\sigma^{\pm}$ transitions. All of these mentioned transition rules are valid for free space but surface plasmon polariton waves obey the inversion of these selection rules [33].

As displayed schematically in Figure (1b, c), we assume that the BSW propagates along the z-axis and the amplitude of its complex electric fields is written as:

$$\vec{E} = (\hat{x} - i\frac{\eta}{k_B}\hat{z}) \exp(ik_B z - \eta x)$$

The π/2 phase difference of longitudinal ($E_z = \frac{\eta}{k_B}$) and transverse ($E_x = 1$) components of the electric Bloch surface wave field implies effective elliptical polarization in the xz plane. In this equation, $k_B$ is the longitudinal wave number (momentum), $\eta$ denotes the transverse wave number (exponential decay rate) and so complex wavevector of BSW is $\vec{k} = \vec{k_B}\hat{z} + i\vec{\eta}\hat{x}$. For evanescent waves, the momentum of an evanescent wave ($k_B$) is perpendicular to its decay direction $\eta$. Also, to satisfy the transverse conditions (K. E=0), the two components of wave vector must have a phase difference of 90 degrees. Universal basic vectors for evanescent waves are expressed as a right-handed triplet consisting of k$_B$ (momentum), $\eta$ (decay) and $k_B \times \boldsymbol{\eta}$ (transverse spin) which controls the direction of lateral forces. Spin-momentum locking in which the direction of momentum is locked into the direction of an intrinsic transverse spin, is an inherent feature of evanescent electromagnetic waves, including the Bloch surface wave [11].

Thus, according to this universal basis vector, we re-define the Faraday and Voigt geometry conditions for evanescent waves as follows: the Faraday configuration, in which the direction of the transverse spin, S, and an external magnetic field, B, are parallel (and perpendicular to the polarization ellipse), induces selection rules of $\Delta m_F = \pm 1$ (σ+ and σ-, respectively) (Figure 1b). In Voigt configuration, in which the magnetic field is transverse to the direction of the spin, selection rules are $\Delta m_F = 0, \pm 1$ (transitions are coined π transitions and σ+ and σ-, respectively). Note that the external magnetic field is not necessarily perpendicular to the electric field components of the Bloch surface wave (Figure1c).

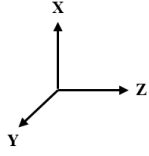
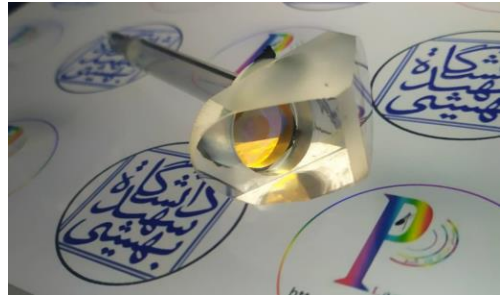
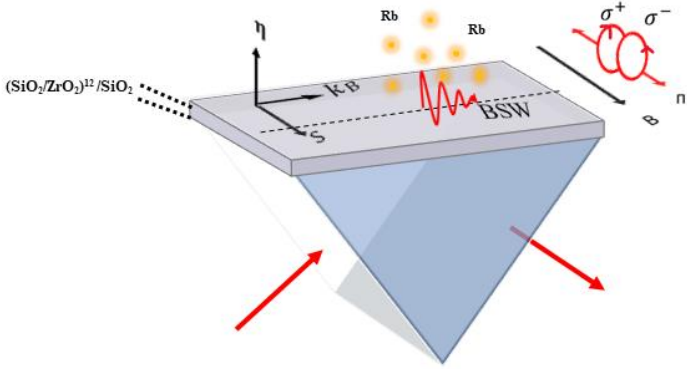
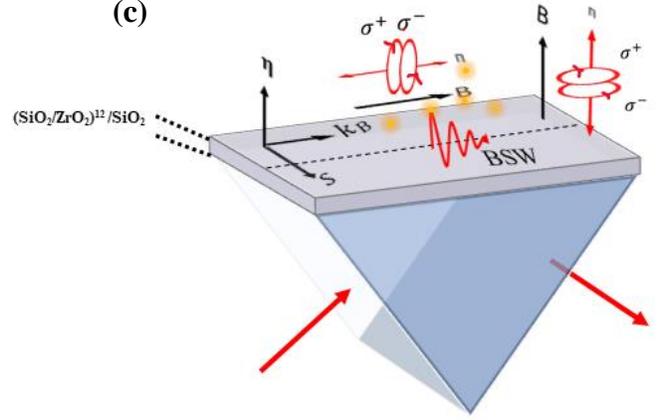

*Figure1: Faraday and Voigt geometry conditions for evanescent wave such Bloch surface waves: (a) photograph of the hybrid BSW-atomic cell, (b) Faraday condition for evanescent waves; schematics of the BSW-atomic system when the magnetic field is applied parallel to the s direction, (c) Voigt condition for evanescent waves; schematics of the BSW-atomic system when the magnetic field is applied transverse to the s direction.*

We measured the ratio of electromagnetic field polarization components by specifying the ratio of the $\sigma^\pm$ transitions. Thus, we can express the elliptical polarization of BSW in right-handed, $|R\rangle$, and left-handed, $|L\rangle$, circular polarization basis as follows:

$$\vec{E} = |L\rangle\langle L|E\rangle + |R\rangle\langle R|E\rangle \quad (1)$$

$$\vec{E} = 1/\sqrt{2}(1+\frac{\eta}{k_B})|R\rangle + 1/\sqrt{2}(1-\frac{\eta}{k_B})|L\rangle \quad (2)$$

According to this analysis, the ratio $\frac{1+\eta/k_B}{1-\eta/k_B}$ is given by the ratio between R and L basis. Therefore, by measuring the transmission spectrum of the reference cell in the Faraday configuration using right/left-circularly polarized light (RCP and LCP) (under the same magnetic field), we can determine the contributions of the right-handed and left-handed circular transitions, respectively.

### III. Results and Discussion:

By adjusting the incident light angle to the prism at the BSW angle we measured the reflection spectra of the BSW-atomic hybrid system in the presence of the external magnetic field. In the first step, according to Figure 1(a), we apply a magnetic field to the hybrid system parallel to the direction of the spin (Faraday configuration for evanescent waves). The measured reflection spectrum from the interface of the photonic crystal and the atomic vapor is represented by the red spectrum in Figure 2(c).

As mentioned above, in the Faraday configuration for evanescent waves ($B \parallel S$) we expect only transitions with $\Delta m_F = \pm 1$ are allowed. To prove this assumption and to determine the contribution of right and left-handed circular polarization to the measured spectrum, as depicted in Figure 2(a, b), we measured the transmission spectra of the reference cell in the cases of right and left-handed circularly polarized light propagating in atomic media where the applied magnetic field was parallel to the wave vector of light (free space Faraday configuration). As stated in the previous section, according to selection rules, the left and right circular polarized light (with $m_j = \pm 1$ acting as angular momentum eigenstates of photons) excite σ − and σ + transitions, respectively. Note that in our frequency modulation spectroscopy setup, the measured spectrum of the hybrid system is an anomalous dispersion-like line shape caused by BSW-atom coupling while the measured spectrum of the reference cell is a normal dispersion-like line shape. Thus, for a better comparison of these two spectra, the measurement spectrum of the hybrid system has been inverted.

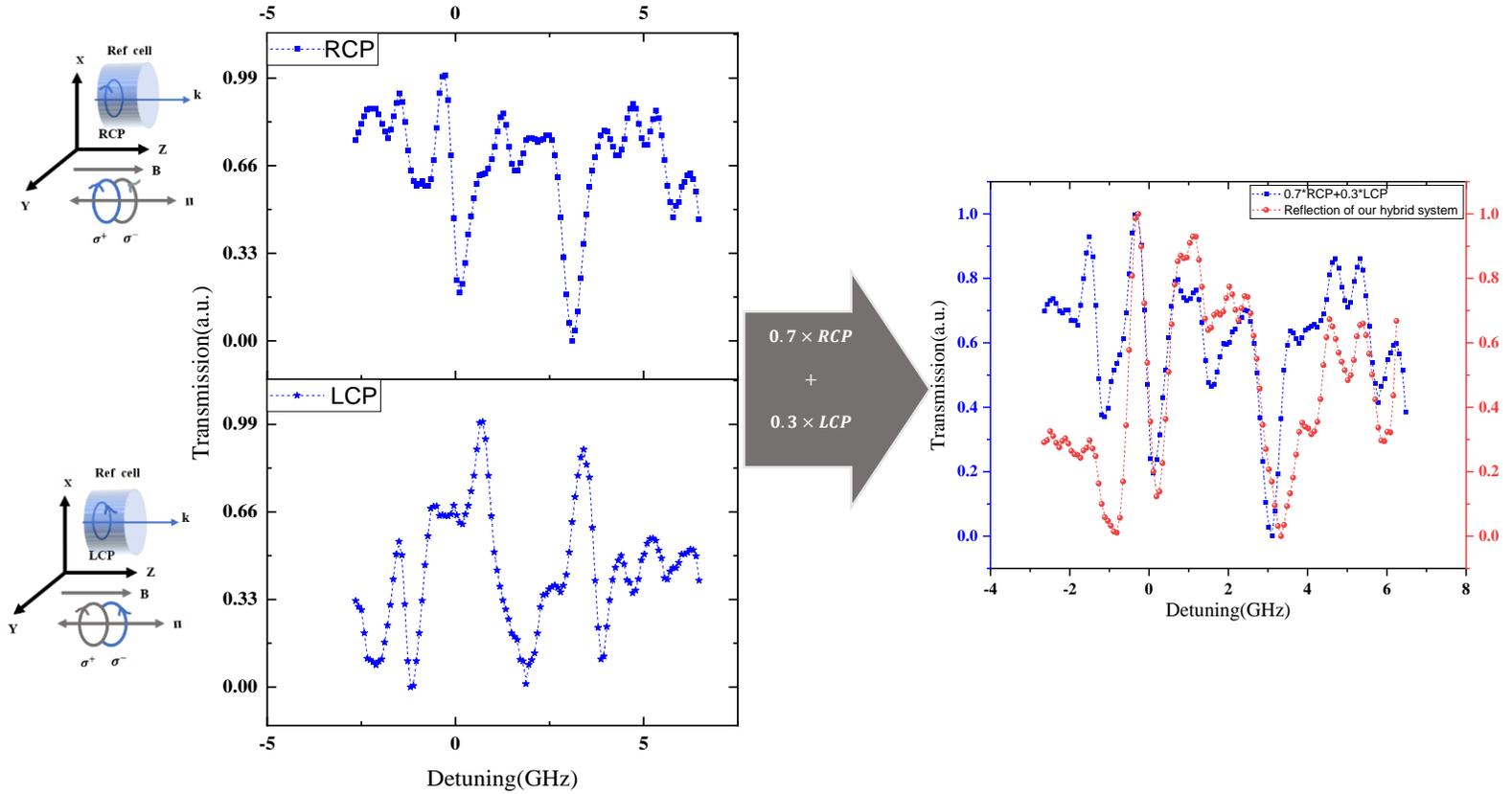

*Figure (2): (a), (b) Transmission spectrums of reference cell for RCP and LCP light, respectively. The magnetic field is applied in free space Faraday configuration. The schematic of these measurements is shown on the left side of these spectrums, (c) Spectrum of hybrid BSW-atomic structure (red curve) when magnetic field is applied parallel to the s direction (Faraday configuration for evanescent waves) and normalized spectra of summed right- and left-handed circular transitions with ratios of 7:3(blue curve).*

As evidenced in Figure 2, we summed the normalized spectra of right- and left-handed circular transitions with ratios of 7:3. Clearly, we find high accordance between 0.7*RCP+0.3*LCP signal (blue line in Figure 2(a, b)) and the measured signal of the hybrid system (red line in Figure 2(c)). The inconsiderable difference between the two graphs is caused by the logical difference between the measurement and the ideal percentage of LCP and RCP light which must be used. Therefore, the ratio between left-handed and right-handed circular polarization is obtained as 7:3, which is equal to $\frac{1+\eta/k_B}{1-\eta/k_B}$, according to Equation 2. Now, we obtained the ratio between the BSWs electric field components and BSW wavenumbers in the x and z directions as follows:

$$\frac{E_z}{E_x} = \frac{k_x}{k_z} = \eta/k_B = \frac{2}{5}$$

This means the Bloch surface wave electric field with elliptical polarization rotates in the xz plane while the ratio of its longitudinal and transverse electric field components equal to 2/5.

In the second step, we measured the reflection spectrum of our hybrid system in the Voigt configuration for evanescent fields according to the schematic of Figure 1(c) as shown in Figure 3(a). In Voigt configuration, the magnetic field can be applied in the direction of BSW propagation or the direction of BSW decay. As mentioned in the previous section, in this geometry we expect that three transitions ($\Delta m_F = 0, \pm 1$) are allowed because the longitudinal and transverse components of the BSW electric field excite $\pi$ and the $\sigma^\pm$ transitions respectively.

As shown in Figure 3, we were able to behold such an effect. To express this observation more precisely, we first measured free space transmission spectrum of the reference cell by incoming TE Linear polarized laser light in Voigt configuration. As the schematic of Figure 3(c) shows, the incident TE polarized light only excites the $\pi$ atomic transitions. Therefore, the magnetically affected spectrum for reference cell only includes $\pi$ transitions. also Figures (3b, d) represent measured transmission spectra which include $\sigma^+$ and $\sigma^-$ transitions respectively.

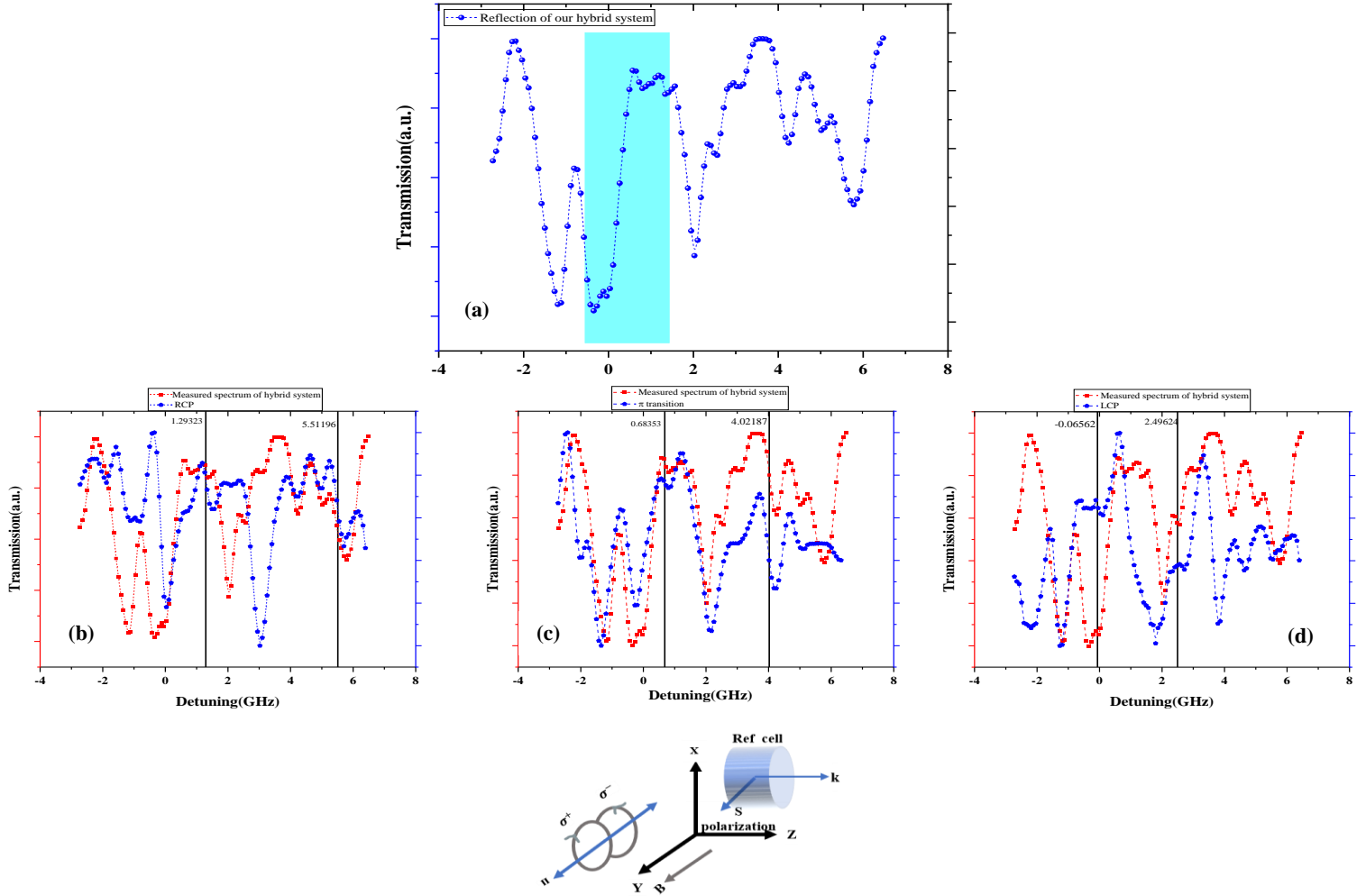

*Figure (3): (a) Measured spectrum of the hybrid BSW-atomic system in Voigt configuration for evanescent waves. The magnetic field is applied perpendicular to the spin direction, (b), (c), (d) Comparison of the transmission spectra including $\sigma^+$, $\pi$ and $\sigma^-$ transitions respectively with the spectrum measured from the hybrid BSW-atomic system. The vertical lines in each figure indicate the corresponding transitions in the two spectra.*

The method of measuring the spectrum of $\sigma^{\pm}$ transitions was mentioned in the previous section. Next, we compared the magneto-optical response of the hybrid system with each of these spectra, which include $\sigma^+$, $\pi$ and $\sigma^-$ transitions respectively. In Figure 3a, the blue highlighted part of the spectrum shows all three transitions in this hybrid system are allowed.

## IV. Conclusion:

In summary, we have proposed a scheme to investigate the polarization of Bloch surface waves using the coupling of BSW and hot atoms. For this purpose, we have studied the coupling of BSW and the hot vapor of Rb in the presence of an external magnetic field. As the first result of these studies, considering the intrinsic property of damped waves, spin-momentum locking, we redefined the Faraday and Voigt configuration based on the orientation of the external magnetic field and the inherent transverse spin of evanescent waves, especially BSW. By evanescent spectroscopic of the BSW-atomic hybrid system in the Faraday configuration of the evanescent wave, we have shown the ellipticity of Bloch surface wave polarization and measured the ratio between the BSW electric field transverse and longitude components. As a result, this method is appealing toward mapping the electromagnetic polarization components and near-field vectorial imaging which is an important tool that enables the visualization and analysis of the vector nature of electromagnetic fields at nanometer scales, opening up a wide range of scientific and technological possibilities.

**Conflict of interest:**

There is no any conflict of interest.

**Data availability:**

The data use in this manuscript can be available by request.